\documentclass[12pt, letterpaper]{article}

\usepackage[utf8]{inputenc}
\usepackage{times}
\usepackage{changes}
\usepackage{color}
\usepackage{amsfonts}
\usepackage{amsmath,amssymb}
\usepackage[english]{babel}
\usepackage{subfigure}
\usepackage{graphicx}
\usepackage{epsf} 
\usepackage{csquotes}
\usepackage{authblk}
\usepackage{hyperref}
\usepackage{gensymb}
\usepackage{scicite}
\usepackage{xfrac}

\newcommand{\ket}[1]{\left|#1\right\rangle}

\newcommand{\braket}[2]{\left\langle #1|#2\right\rangle}

\newcommand{\rom}[1]{\uppercase\expandafter{\romannumeral #1\relax}}

\usepackage{array}
\newcommand{\PreserveBackslash}[1]{\let\temp=\\#1\let\\=\temp}
\newcolumntype{C}[1]{>{\PreserveBackslash\centering}p{#1}}

\begin{document}

\title{Quantum simulations of interacting systems with broken time-reversal symmetry }
\author{Yotam Shapira$^{1, \ast, \dagger}$} 
\author{Tom Manovitz$^{1,\ast, \dagger, \ddagger}$}
\author{Nitzan Akerman$^1$}
\author{Ady Stern$^2$}
\author{Roee Ozeri$^1$}

\affil{\small{$^1$Department of Physics of Complex Systems\\
$^2$Department of Condensed Matter Physics\\
Weizmann Institute of Science, Rehovot 7610001, Israel\\ 
$^*$ These authors contributed equally to this work\\
$^\ddagger$ \emph{Current affiliation:} Department of Physics, Harvard University, Cambridge, MA 02138, USA \\
$^\dagger$ Corresponding authors
}}

\maketitle

\begin{abstract}
    Many-body systems of quantum interacting particles in which time-reversal symmetry is broken give rise to a variety of rich collective behaviors, and are therefore a major target of research in modern physics \cite{halperin2020fractional}. Quantum simulators can potentially be used to explore and understand such systems, which are often beyond the computational reach of classical simulation \cite{cooper2020fractional,altman2021quantum}. Of these, platforms with universal quantum control can experimentally access a wide range of physical properties. However, simultaneously achieving strong programmable interactions, strong time-reversal symmetry breaking, and high fidelity quantum control in a scalable manner is challenging. Here we realized quantum simulations of interacting, time-reversal broken quantum systems in a universal trapped-ion quantum processor \cite{manovitz2021trapped}. Using a scalable scheme that was recently proposed \cite{manovitz2020quantum} we implemented time-reversal breaking synthetic gauge fields, shown for the first time in a trapped ion chain, along with unique coupling geometries, potentially extendable to simulation of multi dimensional systems. Our high fidelity single-site resolution in control and measurement, along with highly programmable interactions, allow us to perform full state tomography of a ground state showcasing persistent current, and to observe dynamics of a time-reversal broken system with nontrivial interactions. Our results open a path towards simulation of time-reversal broken many-body systems with a wide range of features and coupling geometries.
\end{abstract}

\section*{Introduction}
Quantum many-body systems are abundant in nature and lie at the heart of contemporary physical research. These systems exhibit complex collective behaviour and interesting phases which emerge due to interactions between the systems' constituents. Often, richer phenomena are observed when placing such systems in the presence of time-reversal symmetry-breaking (TRSB) fields, such as magnetic fields. A stark example is the appearance of anyonic excitations in the fractional quantum Hall effect \cite{halperin1984statistics,arovas1984fractional,bartolomei2020fractional,nakamura2020direct}.

Typically, quantum many-body systems are analytically intractable and are not amenable to numerical simulation techniques beyond small sizes. Thus, in the past three decades, tremendous effort has been put forth to develop and employ quantum simulation in order to study such systems \cite{monroe2019programmable,altman2021quantum}. Of the variety of physical platforms used for simulating quantum systems with TRSB, ultracold gases of neutral atom have emerged as particularly prolific \cite{cooper2020fractional}. Here, TRSB magnetic fields can be generated by a variety of techniques, including rotation of the gas, optical dressing or Floquet engineering \cite{cooper2020fractional,aidelsburger2018artificial} and have been utilized to great effect in various experiments \cite{jaksch2003creation,lin2009synthetic,lin2011spin,dalibard2011colloquium,bloch2012quantum,struck2013engineering,aidelsburger2013realization,miyake2013realizing,mancini2015observation,stuhl2015visualizing,aidelsburger2018artificial,cooper2019topological}. Recently, TRSB has also been realized in ultracold gasses trapped in tweezer arrays, along with programmable interactions \cite{periwal2021programmable}.

TRSB quantum simulation has also been realized in other platforms, including qubit systems. Some of these platforms can also serve as fully programmable, universal quantum computers. Progress in such platforms has culminated in TRSB demonstrations with superconducting circuits \cite{roushan2017chiral,neill2021accurately} and neutral atoms held in optical tweezers \cite{lienhard2020realization}. Despite the restricted Hilbert spaces, qubit platforms with TRSB can support a variety of exotic phases such as spin liquids and fractional quantum Hall states \cite{wang2011fractional,bauer2014chiral}, and can make use of a powerful quantum control toolset, potentially enabling state tomography; energy spectrum measurement \cite{neill2021accurately}; measurement of topological string operators \cite{Semeghini2021}; quantum-classical variational optimization \cite{cerezo2021variational}; the use of randomized measurements to estimate a range of complex observables \cite{brydges2019probing,huang2020predicting}; and the measurement of entanglement entropy through swap tests \cite{bluvstein2021quantum}.

Trapped ion chains are one of the leading technologies for both quantum simulation \cite{monroe2019programmable} and quantum information processing (QIP) \cite{postler2021demonstration,pino2021demonstration,egan2021fault}, due to unparalleled quantum control fidelity and measurement accuracy. In a typical trapped ion chain platform, an array of ions is confined to a one dimensional chain in a linear Paul trap, and a qubit is encoded on each ion by isolating two energy levels. Qubits can then be coupled by mediating interaction through the motional modes of the ion chain \cite{sorensen2000entanglement}. A unique advantage of trapped ion platforms for both quantum simulation and QIP is their natural long range connectivity, itself a product of the long-range Coulomb interaction mediating the coupling of ions. These properties have been used in an assortment of state-of-the-art quantum simulations \cite{kokail2019self,joshi2021observing,tan2021domain,kyprianidis2021observation,qiao2022observing}. 

Here we realize an analog quantum simulator of TRSB systems in a small trapped ion quantum processor \cite{manovitz2021trapped}. We use a recently proposed method \cite{manovitz2020quantum}, in which experimentally simple additions to the typical ion QIP toolset provide scalable access to highly programmable interactions, including TRSB synthetic fields and emergent coupling geometries that significantly differ from the one-dimensional physical geometry of the ion chain. Using these flexible tools, as well as the quantum control afforded by our quantum processor, we generate a ground state of an interacting Aharonov-Bohm (AB) ring, which we investigate using state tomography and via ground state certification \cite{hangleiter2017direct}; we observe the gauge-field dependent dynamics of the ring model; and we explore the dynamics of a TRSB triangular ladder system in which excitations interact in a nontrivial way. The latter model may be scaled to simulate two-dimensional systems using the one-dimensional chain. Our results are the first realization of TRSB in a trapped ion chain quantum simulator, and delineate a path towards realizing highly flexible and multidimensional coupling geometries in these systems. Moreover, we observe nontrivial quasiparticle interaction in these contexts.

\begin{figure}
\centering
	\includegraphics[width=1\linewidth]{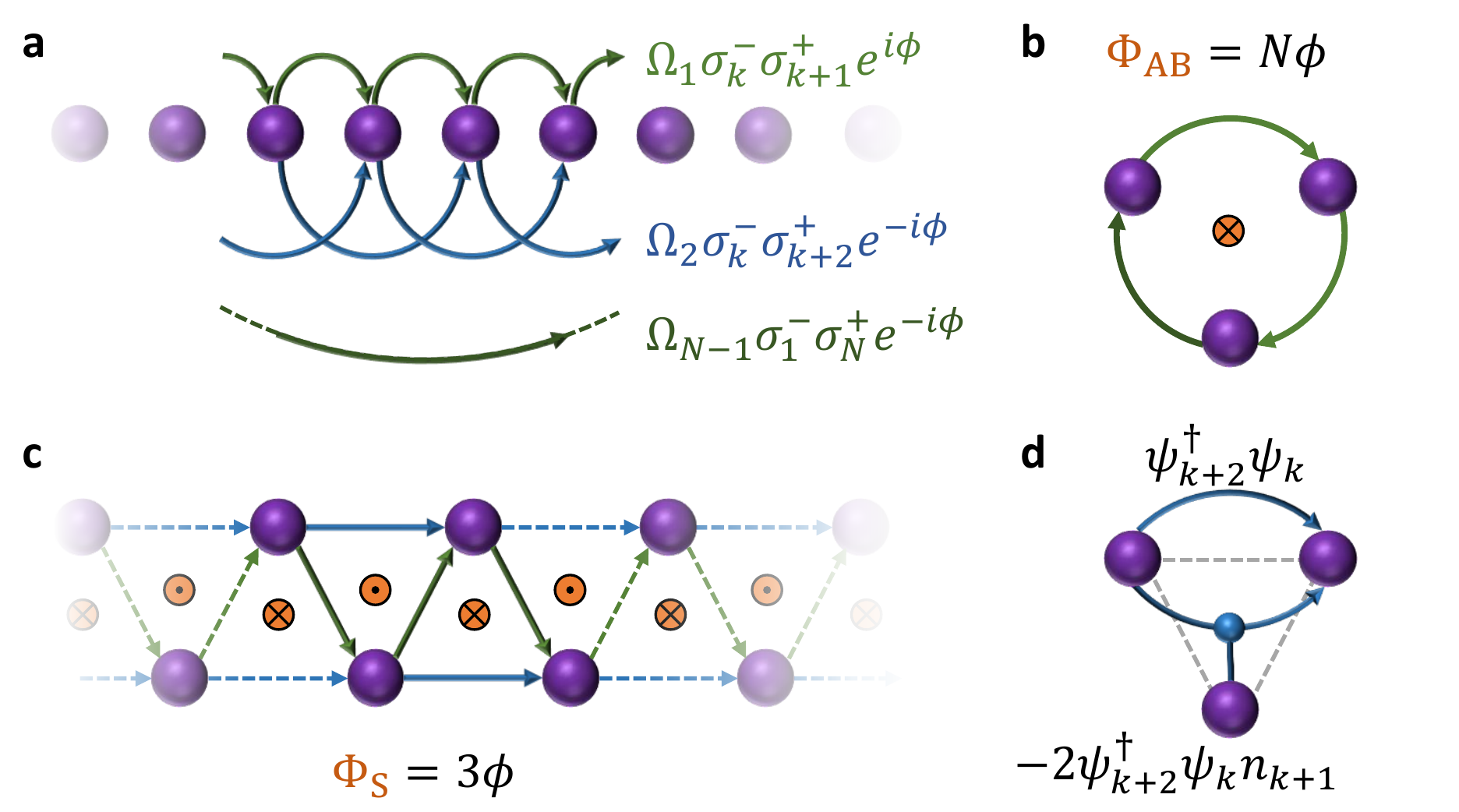}
	\caption{Mapping of an ion chain to spin models. (a) A linear chain of ions is coupled with nn hoping (light-green), nnn hoping (blue) and periodic boundary conditions hoping (dark-green). Each of these interactions is accompanied by a tunable phase, $\phi$. (b) Setting $\Omega_1=\Omega_{N-1}=\Omega$ (and all others to zero) forms an $N$-site AB ring, penetrated by a flux $\Phi_{AB}=N\phi$ (orange). (c) Instead, setting $\Omega_1=\Omega_2=\Omega$ (and all others to zero) forms a triangular ladder. Each plaquette of the ladder is penetrated by a staggered flux, $\Phi_S=3\phi$ (orange). (d) Transforming to a fermionic Hamiltonian reveals that the triangular ladder is an interacting model. Indeed a hop along the ladder's rails is possible through two terms, a 2-body trivial term and an interacting 4-body term for which the hop from site $k$ to $k+2$ is conditioned on the occupation of site $k+1$.}\label{figSystem}
\end{figure}

Our method enables simulation of a variety of models that can be described through the general spin Hamiltonian:
\begin{equation} \label{eqHgeneral}
    H=\sum_{n=1}^{N-1}H_n=\sum_{n=1}^{N-1}\Omega_n e^{i\left(\phi_n-\delta_n t\right)}\sum_{k=1}^{N-n}\sigma_{k+n}^+\sigma_{k}^-+H.c.,
\end{equation}
where $\sigma_k^+$ ($\sigma_k^-$) denotes a spin raising (lowering) Pauli operator on site $k$ out of $N$, and $\Omega_n$, $\phi_n$ and $\delta_n$ are experimentally tunable parameters that respectively correspond to the coupling strengths, static phases and time-dependent phases of an $n$-neighbour hopping interaction. Figure \ref{figSystem}(a) shows a linear chain of ions and couplings utilizing nearest neighbour (nn) with $\Omega_1$, next nearest neighbour (nnn) with $\Omega_2$ and periodic boundary conditions with $\Omega_{N-1}$ .

As we have shown, this simple form encodes a rich family of spin models in various dimensions and geometries. Figure \ref{figSystem} shows the models implemented here, specifically a 3-ion AB ring (b) threaded by a flux $\Phi_{AB}$, and a triangular spin ladder (c), threaded by a staggered flux, $\Phi_S$.

The spin Hamiltonian \eqref{eqHgeneral} can be mapped to a hard-core boson model by identifying a spin excitation with a bosonic excitation $(\sigma^+\leftrightarrow b^\dagger)$, or to a fermionic Hamiltonian through one of several mappings. Indeed, using a Jordan-Wigner transformation \cite{shankar2017exact} the Hamiltonian in Eq. \eqref{eqHgeneral} is rewritten as,
\begin{equation} \label{eqHgeneralFermion}
    H=\sum_{n=1}^{N-1}H_n=\sum_{n=1}^{N-1}\Omega_n e^{i\left(\phi_n-\delta_n t\right)}\sum_{i=1}^{N-n}\psi_{i+n}^\dagger\psi_{i}e^{i\pi\sum_{k=i+1}^{i+n}n_k}+H.c.,
\end{equation}
where $\psi_i$ denotes a fermionic annihilation operator at site $i$ and $n_k=\psi_k^\dagger\psi_k$ is the occupation at site $k$. Using this Hamiltonian form for the triangular spin ladder reveals that the model is interacting, as hopping operators along the ladder's rails are composed of 2-body hop terms and 4-body correlated hop terms, shown in Fig \ref{figSystem}(d).

We note that the spin Hamiltonian \eqref{eqHgeneral} commutes with the total spin (excitation) operator $\sum_{k=1}^N \sigma_k^z$, i.e. the dynamics preserves the total spin, and can therefore be decomposed to total spin excitation subspaces. We utilize this by initializing our system in either the single excitation subspace (1ES) or the two excitations subspace (2ES), which are free of dephasing due to global phase noise. This allows us to observe coherent dynamics for times exceeding our single-qubit coherence time by an order of magnitude \cite{shaniv2018toward}. 

Spin states, $\ket{0}$ and $\ket{1}$, are encoded on the $\ket{5S_\frac{1}{2},m=\frac{1}{2}}$ and $\ket{4D_\frac{5}{2},m=\frac{3}{2}}$ orbitals, respectively, of $^{88}\text{Sr}^+$ ions held in a linear Paul trap. The ions are coherently controlled along this transition using a narrow-linewidth 674 nm laser \cite{peleg2019phase}, addressing the ion chain through a global path, which homogeneously illuminates all ions, and an individual path, in which a tightly focused beam can address a single ion at a time. In order to apply Hamiltonians of the form \eqref{eqHgeneral}, it is necessary to form a spin energy gradient across the chain, and to drive the system with a polychromatic field \cite{manovitz2020quantum}. We realize the former using a magnetic field gradient along the ion chain axis, generating an approximately 1 kHz frequency difference between neighboring ions. A polychromatic field is generated by driving an acousto-optic modulator controlling the global 674 nm path using an arbitrary waveform generator, resulting in effective couplings, $\Omega_n$, of few 100s of Hz. Our implementation is amenable to scaling up as the addition of more sites to the models requires only slight modifications to the spectral components driving the interaction \cite{manovitz2020quantum}. 

\section*{Time reversal symmetry breaking in the Aharonov-Bohm ring}
The AB ring consists of $N$ sites placed on a ring, with nearest-neighbor hopping allowed, and threaded by a magnetic flux $\Phi$. Due to the AB effect, the system exhibits persistent currents, which survive even in the presence of impurities \cite{butiker1983josephson,cheung1988persistent}. We realize an $N=3$ AB ring and investigate the dynamics of one and two excitations on it. We also adiabatically prepare the ring's ground state in the singly-excited manifold. While the AB ring is an analytically solvable model that has been subject to extensive research \cite{aharonov1959,viefers2004quantum,antonio2013transport,roushan2017chiral,downing2021non}, it is useful for demonstrating a broken time-reversal symmetry, and instructive as a validation of our method.

The AB ring Hamiltonian is given by,
\begin{equation}
    H_{\text{AB}}(\Phi)=\Omega\sum_{n=1}^N\sigma_{i+1}^+\sigma_i^-e^{i\Phi/N}+H.c,\label{eqHAB}
\end{equation}
with $N+1\equiv1$ and $\Omega$ an effective coupling. In order to maximize $\Omega$ we slightly deviate from the adiabatic regime prescribed by Ref. \cite{manovitz2020quantum}, leading to additional unwanted couplings. Thus in our implementation below we generate the modified Hamiltonian, $H_{\text{eff}}=H_{\text{AB}}\left(\Phi_{\text{AB}}\right)+\epsilon H_{\text{AB}}\left(0\right)$. We extract $\epsilon$ from our data below using a fit, obtaining $\epsilon=0.22$ (see supplemental material for details). 

\begin{figure}
\centering
	\includegraphics[width=1\linewidth]{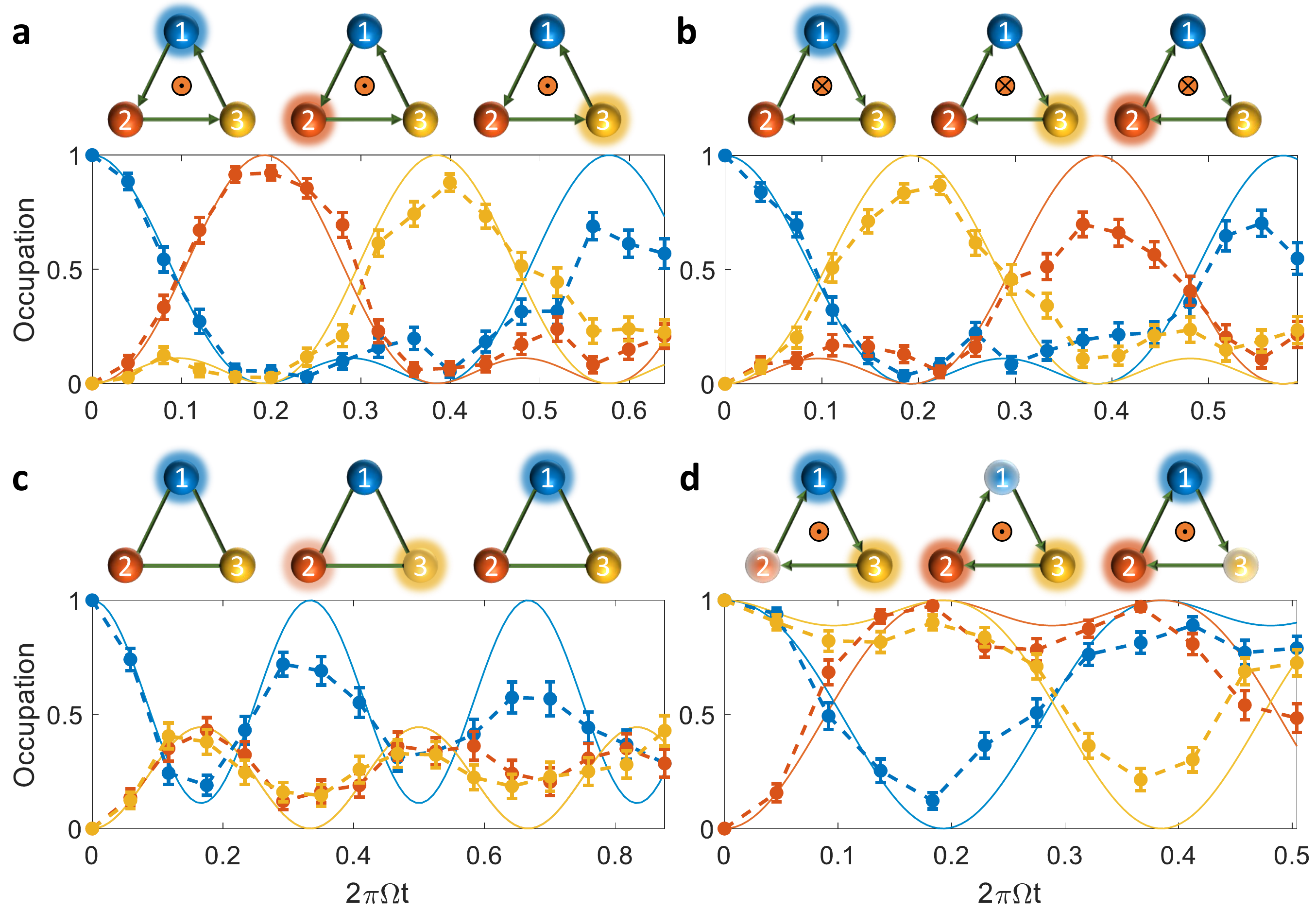}
	\caption{Dynamics in the AB ring, exhibiting TRSB patterns. Schematic (top of each subplot) shows flux direction (orange) and observed spin currents (green arrows). Site colors and numbering correspond to the data (bottom of each subplot). Site occupation data (point and dashed) is post-selected to the relevant excitation subspace and compared to theoretical expectation (solid), showing a good fit. Error bars represent 2$\sigma$ regions due to quantum shot noise. (a) Initialization to site 1, with flux $\Phi_{\text{AB}}=\pi/2$. The excitation 'hops' in a counter-clockwise manner around the ring to sites 2 and then 3, following the spin current arrows. (b) Similarly, by initializing to the same state with $\Phi_{\text{AB}}=-\pi/2$ the evolution becomes clockwise. (c) For $\Phi_{\text{AB}}=0$ time-reversal symmetry is restored and the system does not have a preferred direction of rotation, hence the excitation 'splits' equally clockwise and counter-clockwise and recombines at site 1. (d) Initialization to sites 1 and 2, in the 2ES, with $\Phi=\pi/2$. The system is non-interacting, and can be mapped to a 'hole' (light colors) initialized at site 2.}\label{figABdynamics}
\end{figure}

An apparent manifestation of the persistent currents in this model is transport of excitations around the ring. This is demonstrated by quenching an excitation on an $N=3$ AB ring to specific sites and observing the site occupation at different evolution times. Figure \ref{figABdynamics} shows several examples of these dynamics. The measurements are post-selected, and normalized, to the relevant excitation subspace (see supplemental material). We compare the dynamics to a numerical evaluation of the evolution, with $\Omega$ obtained by a maximum likelihood (ML) fit. 

Figure \ref{figABdynamics}(a) shows the evolution of an excitation on site 1, i.e. the state $\ket{100}$, with $\Phi_{AB}=\pi/2$. Here the excitation traverses the ring sites in a counter-clockwise manner, i.e. $1\rightarrow2\rightarrow3\rightarrow1$. A single excitation in the 3-site AB ring is, to a high degree of accuracy, a wave packet, hence the excitation does not disperse for long evolution times. Moreover, for $\Phi_{\text{AB}}=\pm\pi/2$ the evolution is periodic. (b) Using the opposite flux, $\Phi_{\text{AB}}=-\pi/2$, the evolution becomes clockwise, i.e. $1\rightarrow3\rightarrow2\rightarrow1$. Setting the flux to $\Phi_{\text{AB}}=0$ (c), time-reversal symmetry is restored and the excitation equally occupies site 2 and 3, before recombining back at site 1.

Furthermore, we initialize the system to sites 1 and 2 in the 2ES with $\Phi_{\text{AB}}=\pi/2$, shown in fig. \ref{figABdynamics}(d). This may be transformed to a 'hole' occupying site 3 \cite{roushan2017chiral}, which rotates around the ring in the opposite direction to that of an excitation. This demonstrates that the ring geometry generates a non-interacting model. That is, the dynamics in the 2ES are essentially unmodified and can be deduced from the dynamics in the 1ES. As we show below, this is not the case in the interacting triangular ladder.

Next, we investigate the ground state in the single excitation subspace of the AB ring. This is done by the following steps. We initialize the system in the state $\ket{100}$ in the 1ES. We then turn on the Hamiltonian $H_0=-\delta\sigma_1^z$, which lowers the energy of this state, making it the ground state of $H_0$ in the 1ES. Lastly, we adiabatically ramp down $H_0$ and ramp up $H_{\text{eff}}$. Specifically, the ramp is implemented by the Hamiltonian,
\begin{equation}
    H\left(t\right)=\left[1-\left(t/T\right)^2\right]H_0+(t/T)^2H_{\text{eff}},\label{eqHramp}
\end{equation}
where $T$ is a long ramp time. When adiabaticity holds we expect the resulting state, $\ket{\psi_f}$, to be the ground state of the AB ring in the 1ES subspace. 

\begin{figure}
\centering
	\includegraphics[width=1\linewidth]{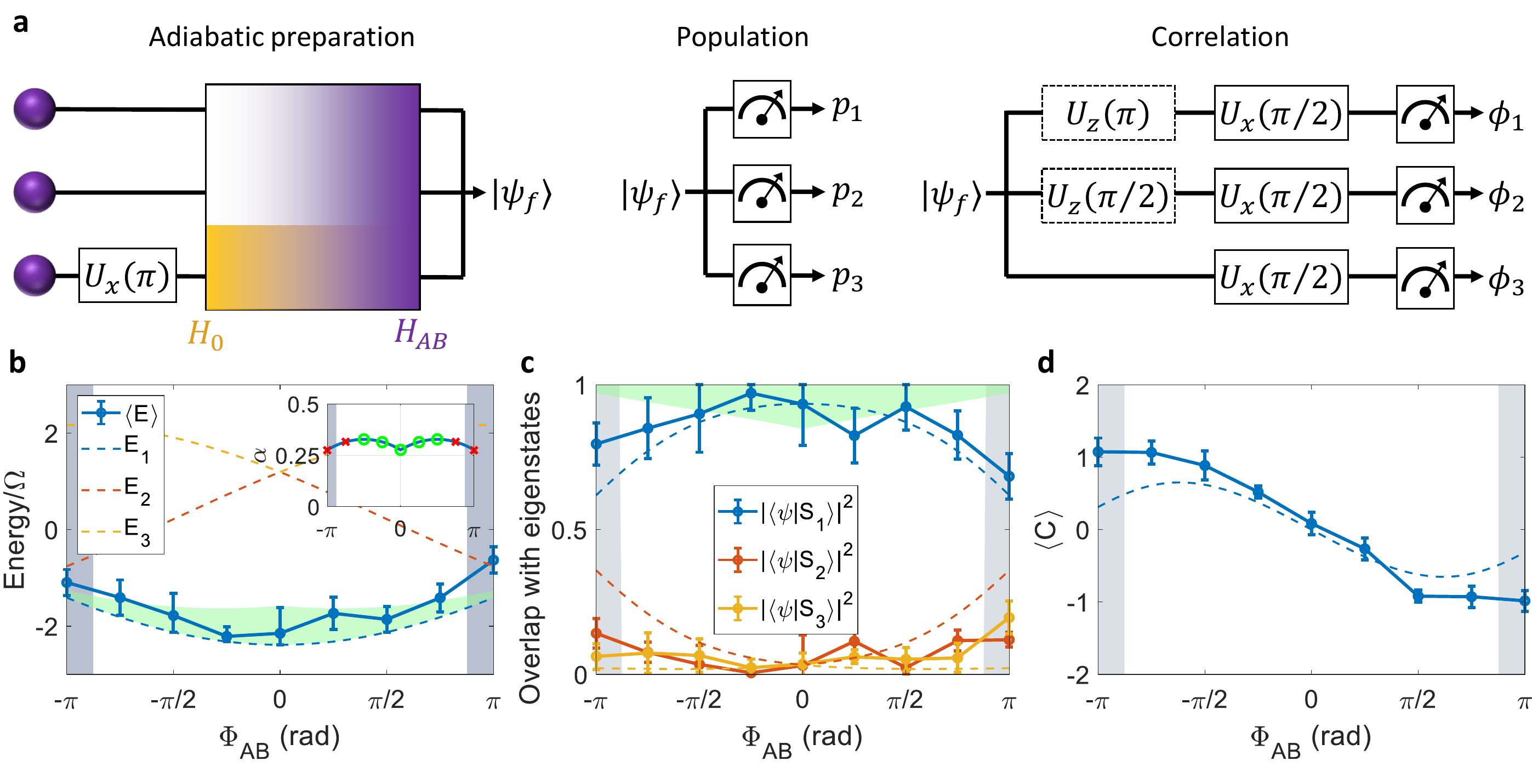}
	\caption{Adiabatic preparation of the AB ring ground state. (a) Preparation and tomography sequence. We initialize the state $\ket{001}$ by rotating a single site with $U_x(\pi)$, lower the site's energy with $H_0$ (orange) and slowly ramp $H_0$ down while ramping $H_{\text{eff}}$ up (purple). The state is prepared at the end of the ramp (left). We preform tomography of our state by measuring site occupation (middle), in-phase correlation via $U_x$ pulses (right) and out-of-phase correlation via additional $U_z$ rotations (dashed). (b) Spectrum of $H_{\text{eff}}$ (dashed) and measured energy of the prepared states (solid), showing good agreement. The spectral gap from the ground state is small at $\Phi_{\text{AB}}=\pm\pi$ leading to a breakdown of adiabaticity, marked by the gray region. 'Accept' energy values for ground state certification are marked in the green region. The inset shows accepted (green) and rejected (red) states and the corresponding reliability, $\alpha$. (c) Overlap of prepared state with the eigenstates, $\ket{\psi_{1,2,3}}$, of $H_{\text{eff}}$. The ground state, $\ket{S_1}$, is accurately prepared in the adiabatic region, with a peak overlap of $0.97\protect\substack{+0.03 \\ -0.05}$. The preparation degrades in the non-adiabatic region. The green region marks the certification fidelity, $F_T+\delta$, for which states are accepted with probability $1-\alpha$. (d) Expectation value of the current operator. The ground state exhibits a persistent non-vanishing current that changes sign around $\Phi_{\text{AB}}=0$. The measured value (solid) is in good agreement with the theoretical expectation (dashed).}\label{figRamp} 
\end{figure}

Utilizing the fact that our quantum simulator is embedded in a universal quantum processor, we preform tomography of the resulting state, $\ket{\psi_f}$. This requires three independent measurements of $\ket{\psi_f}$: occupancy of the ring's sites; in-phase correlation between the sites, obtained by operating with an additional global $x$ spin rotation, $\sum_n U_n^{x}\left(\pi/2\right)$, prior to the measurement; out-of-phase correlation, obtained by preforming an additional, $U_2^{z}\left(\pi/2\right)+U_3^{z}\left(\pi\right)$ rotation, prior to the global $x$ rotation. For a general pure state in the 1ES, $\ket{\psi}=\sqrt{p_1}e^{i\phi_1}\ket{100}+\sqrt{p_2}e^{i\phi_2}\ket{010}+\sqrt{p_3}e^{i\phi_3}\ket{001}$, the occupancy measurement yields the $p_n$'s and the correlation measurements yield the $\phi_n$'s (up to a global phase), which allows us to assess the preparation process in full. Figure \ref{figRamp}(a) schematically shows the adiabatic preparation (left) and measurement sequences (middle and right). We remark that our results use post-selection in the 1ES and assume rapid dephasing of coherences between and in other excitation subspaces.

Clearly, state tomography becomes infeasible for long ion chains, therefore we supplement our analysis with direct certification of ground state preparation using the method in Ref. \cite{hangleiter2017direct}. This method tests ground state preparation of Hamiltonians which are composed of local interactions (e.g. the Hamiltonian in Eq. \eqref{eqHgeneral}), have a known ground state energy and known gap to the next excited state. The method formulates a criterion which is based on local energy measurements of candidate states in order to 'accept' or 'reject' them. States with an overlap with the ground state, $F=\left|\braket{\psi}{\psi_\text{GS}}\right|^2$, which is below a predetermined threshold, $F_T$, will be rejected with high probability, $1-\alpha$. States with a fidelity, $F>F_T+\delta$, will be accepted with high probability, $1-\alpha$, with $\delta$ an additional fidelity gap. No guarantees are given on the acceptance probability of states with fidelity $F_T<F<F_T+\delta$. Here we choose the threshold, $F_T=0.7$.

We repeat this process for various values of flux, $\Phi_{\text{AB}}$, and evaluate the prepared state. Our results are shown in Fig. \ref{figRamp}(b)-(d). The spectrum of $H_{\text{eff}}$ as a function of $\Phi$, is shown in Fig. \ref{figRamp}(b). Energies are normalized by a characteristic coupling scale, $\Omega=350\text{ Hz}$. For $\Phi_{\text{AB}}=0$ the ground state (dashed blue) is well separated from the other states, however at $\Phi_{\text{AB}}=\pm\pi$ the spectral gap is small. Accordingly the gray region (here and in all other subplots) marks fluxes where Landau-Zener transitions limit our ground-state preparation fidelity. Nevertheless, the prepared state's energy (solid blue) fits very well to the ground state energy. Here (and in all other subplots) error bars mark $2\sigma$ errors due to quantum projection noise. 

The green region marks energy values for which state certification \cite{hangleiter2017direct} accepts the prepared ground state. The inset in Fig \ref{figRamp}(b) shows accordingly which states are accepted (green) and rejected (red) and the certification's reliability, $\alpha$. Clearly, when the ground state energy gap is sufficiently large, the prepared states are accepted within a reasonable success rate.

Figure \ref{figRamp}(c) shows the measured (solid) and expected (dashed) overlap of the prepared states with the eigenstates of $H_{\text{eff}}$, based on a simulation (see details in supplemental material). We observe good agreement between measurement and expectation, with a peak overlap with the ground state of $0.97\substack{+0.03 \\ -0.05}$. We further notice that this overlap degrades in the non-adiabatic (gray) regions as expected. The green region marks the fidelity region, $F>F_T+\delta$, of accepted certified states, in correspondence to Fig \ref{figRamp}(b). We note that this region varies with $\Phi_\text{AB}$, as $\delta$ depends on the ground state energy gap.

Finally, our prepared ground state exhibits non-vanishing persistent spin currents, which mark the broken time-reversal symmetry generated by the AB flux. We measured the expectation value of the spin current operator, $C=i\sum_{n=1}^N\left(\sigma_{n+1}^+\sigma_n^-e^{i\Phi_{\text{AB}}/N}-H.c\right)$. As shown in Fig. \ref{figRamp}(d) this expectation value varies between positive and negative values according to $\Phi_{\text{AB}}$, demonstrating good agreement between the theoretical prediction (dashed) and measured (solid) values of spin currents in the ground state.

\section*{Interactions on triangular ladder}
While the one-dimensional Aharonov-Bohm ring is convenient for clearly illustrating TRSB and periodic boundary conditions, it is an exactly solvable free fermion model, and thus is not an interesting target for future quantum simulation. However, by slightly varying the coupling geometry, it is easy to generate models with TRSB which cannot be reduced to free fermions, and which inherently include interactions. 

Specifically we set $\Omega_1=\Omega_2=\Omega$ (and null all other couplings) in the spin Hamiltonian in \eqref{eqHgeneral}, and form triangular plaquettes that make up the ladder coupling configuration shown in Fig. \ref{figSystem}(c). The ladder's rungs (diagonal green lines) are formed by the nearest neighbor term and its rails (horizontal blue lines) are formed by the next nearest neighbor terms. We set $\phi_1=-\phi_2$, thereby generating a staggered flux of $\Phi_S=3\phi$ through the plaquettes. Although the total flux nulls, time reversal symmetry is still broken by  the local gauge invariant flux.

Here we focus on $\Omega_1=\Omega_2=\Omega$; however, $\Omega_1,\Omega_2$ can be fully controlled using our technique, and different coupling strength ratios can generate interesting models which can support a variety of phases, including ordered dimerized, Luttinger liquid, and chiral-ordered phases \cite{majumdar1969next,hikihara2008vector,amico2008entanglement,wang2008tuning,furukawa2010chiral,furukawa2012ground,mishra2013quantum,mishra2014supersolid,anisimovas2016semisynthetic,an2018engineering,brydges2019probing,cabedo2020effective}.

The model's Hamiltonian in its fermionic form is,
\begin{equation}
    H_{\text{tl}}=\Omega\sum_{k}\left(\psi_{k+1}^{\dagger}\psi_{k}+\psi_{k}^{\dagger}\psi_{k+2}\left[1-2n_{k+1}\right]\right)e^{i\phi}+H.c,\label{eqHtlF},
\end{equation}
which is an interacting Hamiltonian. Indeed the  term, $\psi_{n}^{\dagger}\psi_{n+2}$, has a single-particle contribution, but also an interacting 4-body term, which conditions the n.n.n. hop on the occupation of the intermediate site, as depicted in Fig. \ref{figSystem}(d). 

We make use of a gauge transformation and analyze the model in its spin form,
\begin{equation}
    H_{\text{tl}}=\Omega\left(\sum_{k\text{ odd}}\sigma_{k+1}^+\sigma_{k}^-+\sum_{k\text{ even}}e^{i\Phi_S}\sigma_{k+1}^+\sigma_{k}^-+\sum_k\sigma_{k}^+\sigma_{k+2}^-\right)+H.c,\label{eqHtlS}
\end{equation}

The gauge choice in Eq. \eqref{eqHtlS} makes the model's symmetries apparent, which is helpful in its analysis (see detailed model analysis in supplemental material). Namely for a $N=4$ minimal triangular ladder we observe a unitary symmetry, $U_{1,4}=\left(1+\sigma_1^x\sigma_4^x+\sigma_1^y\sigma_4^y+\sigma_1^z\sigma_4^z\right)/2$, which swaps sites $1$ and $4$, an anti-unitary symmetry $A=U_{2,3}K$, with $K$ the complex conjugation operator. Furthermore, the staggered flux, $\Phi_S=\pm\pi/2$ gives rise to an additional chiral symmetry, $C=\sigma_1^z\sigma_4^z U_{2,3}$, which anti-commutes with the Hamiltonian. An analogous fermionic representation can be constructed by the Jordan Wigner transformation.

We first focus on the 1ES. Due to $U_{1,4}$ the state $\ket{S_1}\equiv\ket{S}_{1,4}\ket{00}_{2,3}$, with $\ket{S}_{i,j}$ a singlet of sites $i$ and $j$, is decoupled from the rest of the spectrum, and carries zero energy. Due to the existence of the chiral anti-symmetry $C$ the remaining three states in the 1ES have energies $\pm E_1$ and $0$, where $E_1$ is computed by diagonalization yielding $E_1=\sqrt{5}\Omega$. Thus the 1ES spectrum is equally spaced, leading to a periodic evolution with period $T_{1ES}=2\pi/\sqrt{5}\Omega$. We quench the system by initializing a single local excitation. 

\begin{figure}
\centering
	\includegraphics[width=1\linewidth]{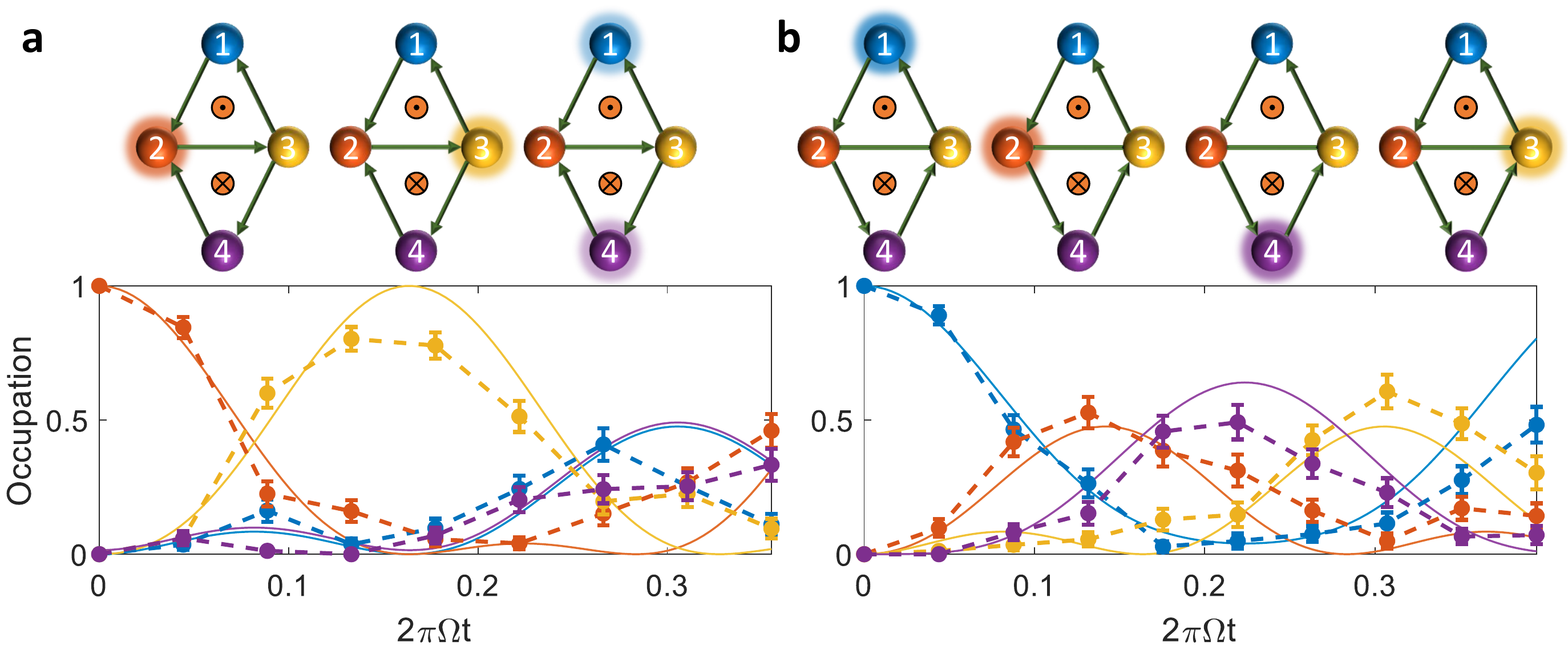}
	\caption{Dynamics on triangular ladder in the 1ES with $\Phi_S=\pi/2$, exhibiting TRSB patterns. Schematic (top) shows staggered flux direction (orange) and observed spin-currents (green arrows). Site colors and numbering correspond to the data (bottom). Site occupation data (point and dashed) is post-selected to the 1ES and compared to theoretical expectation (solid), showing a good fit. Error bars represent 2$\sigma$ regions due to quantum projection noise. (a) Initialization at $\ket{2}$. The evolution follows the green spin-current arrows, i.e the state hops to site 3, then splits to sites 1 and 4 and finally recombines at 2. For clarity, the theoretical blue curve for site 1 has been slightly shifted. (b) Initialization to $\ket{1}$. The peak population encompasses the ladder in a counter-clockwise manner.}\label{figTriangular1ES}
\end{figure}

We initialize the system to $\ket{2}=\sigma_2^+\ket{0000}$ in the 1ES and observe the system's dynamics. This state is an eigenstate of $U_{1,4}$ with an eigenvalue of $+1$, thus this symmetry is preserved throughout the evolution. Indeed Fig. \ref{figTriangular1ES}(a) shows the propagation of the excitation on the ladder schematically (top) and as measured in the 1ES (bottom). Similarly to the AB ring above, the evolution exhibits TRSB and is determined by the local fluxes through the ladder's plaquettes. The excitation hops to site 3, then splits to sites 1 and 4, and finally recombines back at site 2. That is, following the depicted green current arrows. We note that due to symmetry $U_{1,4}$, the  occupation of site $1$ and $4$ can only be in a symmetric triplet sate $\ket{T_1}=\ket{T}_{1,4}\ket{00}_{2,3}$, excluding $\ket{S_1}$ from the dynamics.

Next, we initialize the system to site $\ket{1}$ as shown in Fig. \ref{figTriangular1ES}(b). Similarly to above, TRSB is manifested in the spin-excitation trajectory. We observe the peak occupation encompassing the ladder in a counter-clockwise manner, following the path $1\rightarrow2\rightarrow4\rightarrow3\rightarrow1$. Here the initial state is an equal superposition of $\ket{S_1}$ and $\ket{T_1}$, hence the occupation of site 4 is a result of an accumulation of a phase difference between these states. The spin current patterns demonstrated here are in agreement with Ref. \cite{cabedo2020effective}, for the frustration-free regime. We compare the dynamics shown in Fig. \ref{figTriangular1ES}(a) and (b) to a numerical evaluation of the evolution, with $\Omega$ given by a ML fit, yielding $\Omega=245\pm2\text{ Hz}$ with $95\%$ probability. 

We now turn to the 2ES. Here the four-body correlated hop term, shown in Eq. \eqref{eqHtlF}, becomes active and modifies the system's evolution. A similar symmetry based analysis reveals that the 2ES spectrum is composed of the enrgies: $0$, $\pm\Omega$ and $\pm E_2$; with $E_2$ evaluated by direct diagonalization as $3\Omega$. We conclude that the 2ES evolution as well is periodic with period $T_{2ES}=2\pi/\Omega$. The difference between $E_1$ and $E_2$ is a clear signature of interactions, i.e. the four-body interaction in Eq. \eqref{eqHtlF} modifies the 2ES spectrum (when removing the 4-body term from Eq. \eqref{eqHtlF} the spectrum is $\pm E_1,0$ for all eigenstates).

\begin{figure}
\centering
	\includegraphics[width=1\linewidth]{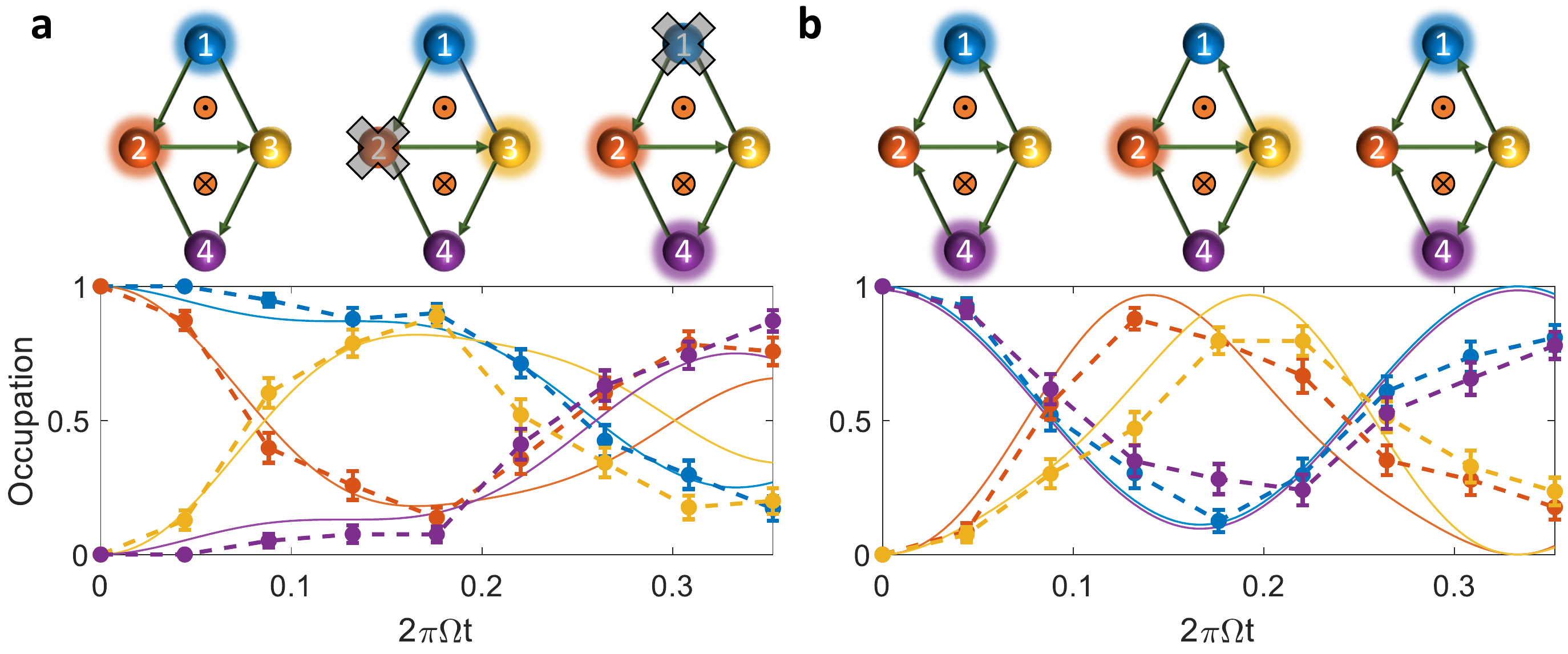}
	\caption{Dynamics on triangular ladder in the 2ES with $\Phi_S=\pi/2$, exhibiting interactions. Schematic (top) shows staggered flux direction (orange) and observed spin-currents (green arrows). Site colors and numbering correspond to the data (bottom). Site occupation data (point and dashed) is post-selected to the 2ES and compared to theoretical expectation (solid), showing a good fit. Error bars represent 2$\sigma$ regions due to quantum shot noise. (a) Initialization in $\ket{1,2}$. Evolution is affected by the four-body interaction term, resulting in a hop to $\ket{1,3}$, as opposed to the 1ES evolution $1\rightarrow2$ (represented by a cross over site 2). The next hop similarly is to $\ket{2,4}$ which does not follow the 1ES evolution of $3\rightarrow1$. (b) Initialization to $\ket{1,4}$. The system  exhibits oscillations with period $T_{2ES}/3$. For clarity, the theoretical curve for site 4 has been slightly shifted.}\label{figTriangular2ES}
\end{figure}

We initialize the system to the state $\ket{1,2}$, with $\ket{i,j}=\sigma_i^+\sigma_j^+\ket{0000}$, as shown in Fig. \ref{figTriangular2ES}(a). This initial state evolution is directly affected by the four-body term, $\psi_1\psi_3^\dagger n_2$, i.e. the presence of an excitation on site 2 modifies the hop $1\rightarrow3$. Indeed we observe modified dynamics with respect to the separate evolution of these two excitations in the 1ES. Specifically the state evolves to $\ket{1,3}$, as opposed to the 1ES evolution $\ket{1}\rightarrow\ket{2}$ in Fig. \ref{figTriangular1ES}(a) and (b). Similarly the state then evolves to $\ket{2,4}$ as opposed to the 1ES evolution $\ket{3}\rightarrow\ket{1}$. 

Next, we initialize the system in the state $\ket{1,4}$, which is a $+1$ eigenstate of $U_{1,4}$, shown in Fig. \ref{figTriangular2ES}(b). The initial state occupies only $E=0$ and $E=\pm 3\Omega$ states, thus the evolution is periodic with period $T_{2SE}/3=2\pi/3\Omega$, which is observed in the data. We compare the dynamics shown in Fig. \ref{figTriangular2ES}(a) and (b) to a numerical evaluation of the evolution, with $\Omega$ given by a ML fit, yielding $\Omega=246\pm2\text{ Hz}$ with $95\%$ probability, in complete agreement with the 1ES data, confirming the theoretical modeling of the system, including the interactions between excitations in the 2ES.

\section*{Conclusion}
Our measurements, shown above, constitute a first demonstration of quantum simulations of TRSB systems in a trapped ion quantum computer, enabled by the methods detailed in \cite{manovitz2020quantum,manovitz2021trapped}. We observe the complex dynamics generated by the combination of strong interactions and TRSB. The flexibility and scalability of our methods will enable future exploration of more complex models, measuring both non-equilibrium and ground state properties of a variety of coupling geometries. The embedding of the simulator in a programmable quantum computer can be leveraged to certify ground states, as shown, but also to measure complex observables, including high order correlations needed to probe topological order parameters;  measure nonlinear observables such as entanglement entropy through the use of randomized measurements; and the use of hybrid quantum-classical optimization to probe ground states complex spin systems.

We thank Rotem Arnon-Friedman, Anna Keselman, Meirav Pinkas, Boaz Raz and Hagai Edri for helpful discussions. This work was supported by the Israeli Science Foundation, the Israeli Ministry of Science Technology and Space, and the Minerva Stiftung.

\bibliographystyle{Science}
\bibliography{references.bib}

\section*{Supplemental material}
Here we provide additional supplementary data to that showed in the main text and additional information regarding data analysis.

\subsection*{Quench dynamics data}

Figures \ref{figABdynamics}, \ref{figTriangular1ES} and \ref{figTriangular2ES} show dynamics of the AB ring and triangular ladder models, quenched to specific sites. The occupation data shown in the figures is normalized to the occupation of the relevant excitation subspace. For example, the date shown in Fig. \ref{figABdynamics}(a) is normalized by the occupation of the single excitation subspace. The spin excitation is in general not conserved due to decoherence and non-adiabatic effect \cite{manovitz2020quantum}. Here we show the evolution of the subspace occupation, corresponding to the data in the main text.

\begin{figure}
\centering
	\includegraphics[width=1\linewidth]{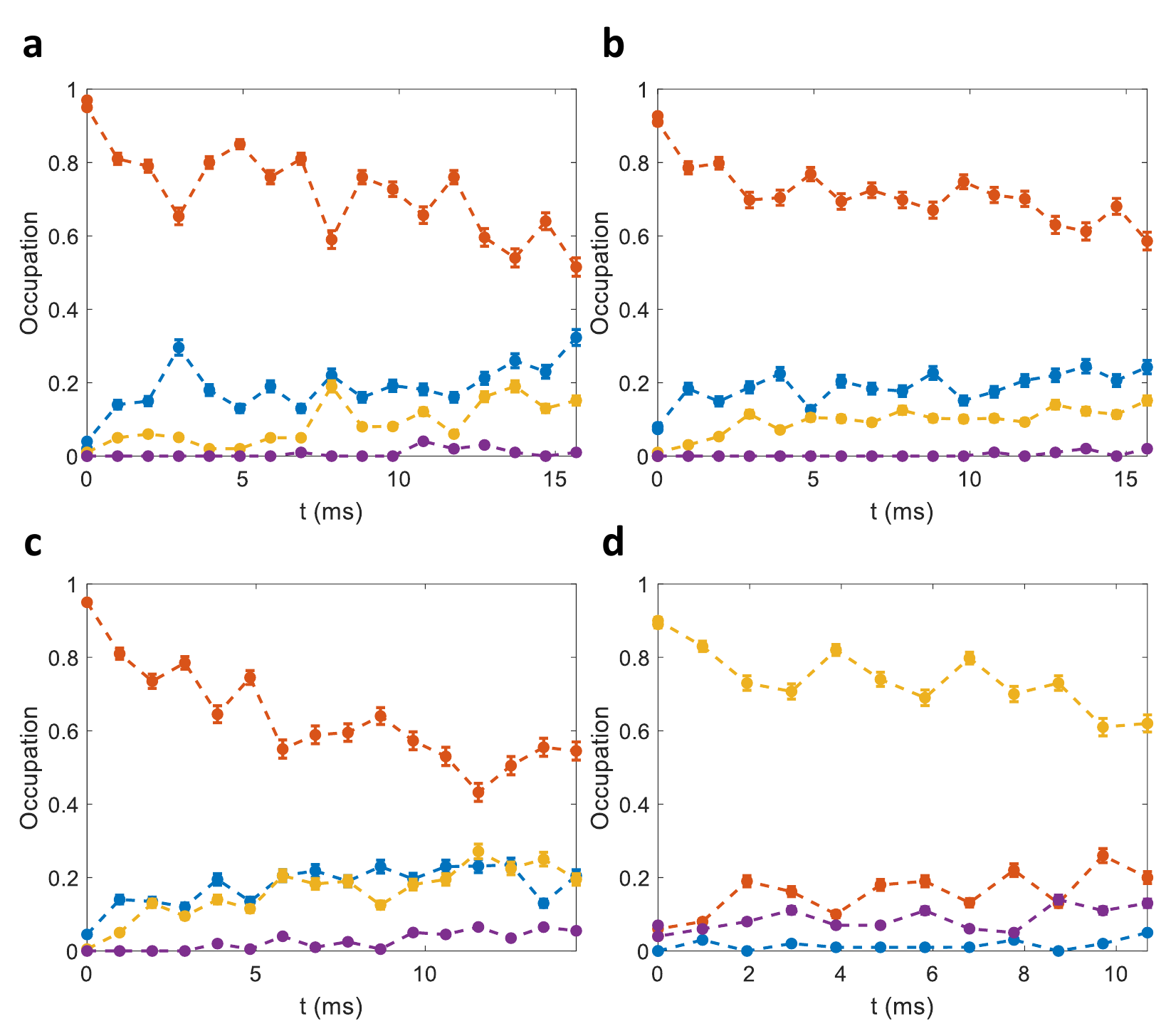}
	\caption{Subspace occupancy of AB ring experiments. With zero (blue), one (red), two (orange) and three (purple) excitations present. (a-d) corresponding to Fig. \ref{figABdynamics}(a-d) of the main text.}\label{figABSubspace}
\end{figure}

Figure \ref{figABSubspace} shows the subspace occupation in the AB ring model, corresponding to Fig. \ref{figABdynamics} of the main text. We note that the initial occupation decays, but remains larger than $0.5$ throughout the evolution.

\begin{figure}
\centering
	\includegraphics[width=1\linewidth]{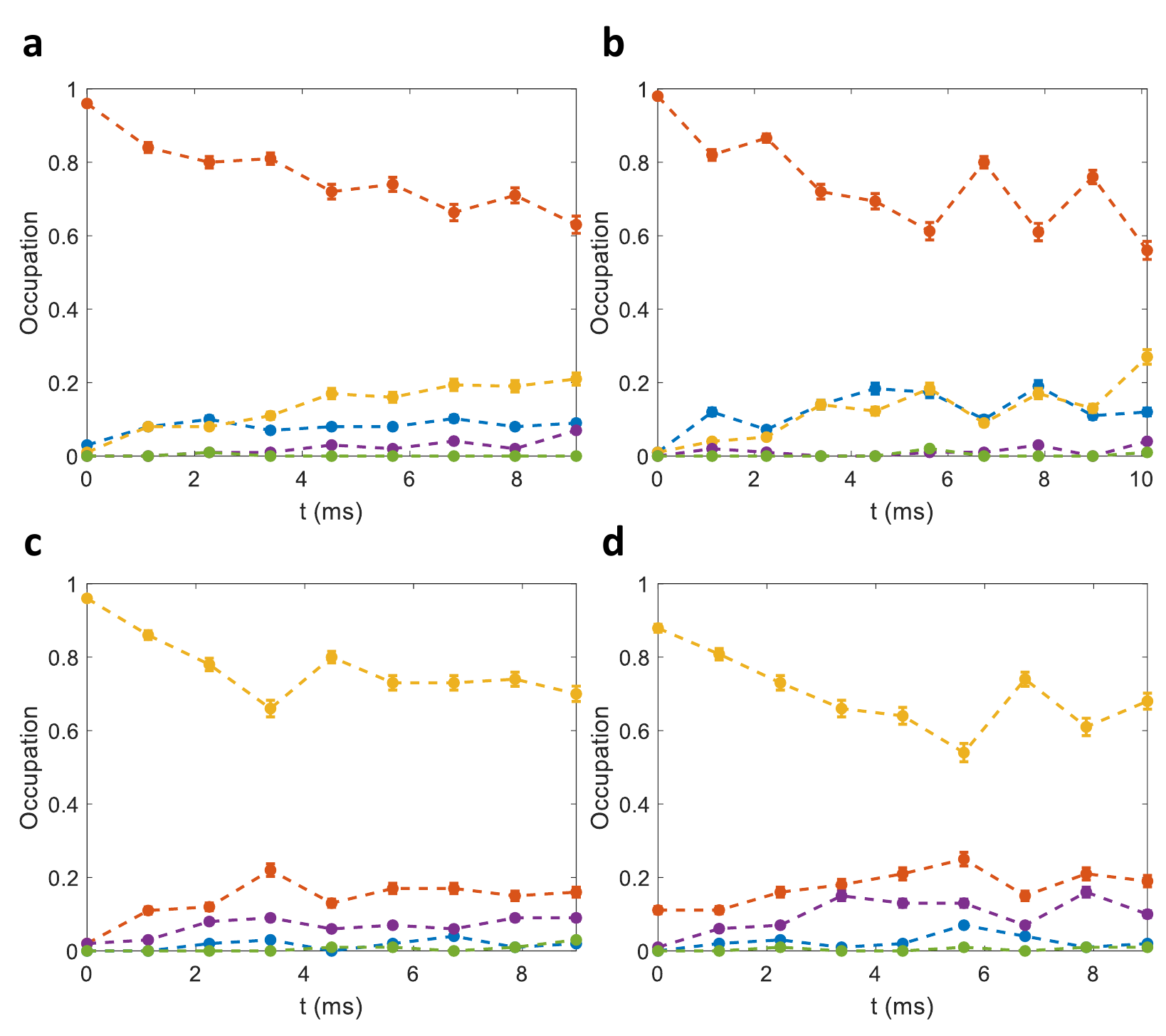}
	\caption{Subspace occupancy of triangular ladder ring experiments. With zero (blue), one (red), two (orange), three (purple) and four (green) excitations present. (a,b) corresponding to Fig. \ref{figTriangular1ES}(a,b) of the main text. (c,d) corresponding to Fig. \ref{figTriangular2ES}(a,b) of the main text.}\label{figTriangularSubspace}
\end{figure}

Figure \ref{figTriangularSubspace} shows the subspace occupation in the triangular ladder, corresponding to Fig. \ref{figTriangular1ES} and \ref{figTriangular2ES} of the main text.

\subsection*{Adiabatic ground state preparation}

In the main text we state that we do not precisely implement the AB Hamiltonian, $H_{\text{AB}}\left(\Phi_\text{AB}\right)$, in Eq. \eqref{eqHAB}. Rather, we implement the Hamiltonian, $H_{\text{eff}}=H_{\text{AB}}\left(\Phi_{\text{AB}}\right)+\epsilon H_{\text{AB}}\left(0\right)$, such that the factor that accompanies a hop is modified, $e^{i\Phi_\text{AB}}\rightarrow e^{i\Phi_\text{AB}}+\epsilon$. This modification can be understood as non-adiabatic corrections to the method in \cite{manovitz2020quantum}. Using this method we generate the spin-hopping terms by utilizing a two-photon process. Specifically the ions are driven by pairs of laser tones, such that absorption of one tone and emission of the other exactly bridges the energy gap to adjacent ions. The phase associated with such a hop is given by the phase difference of these two tones.

However, a similar effect may occur by absorption and emission of photons of the same tone, such that the phase difference vanishes, yielding a vanishing flux. This effect is non-resonant and is characterized by the dimensionless number, $\beta=\xi/2\Omega_n$, where, similarly to power broadening effects, $\xi$ is the detuning from the transition and $\Omega_n$ is the Rabi frequency of the tone. In our implementation we used $\beta=7$, and 8 different tones, yielding the estimate, $\epsilon\approx8/7^2\approx0.16$. 

In practice we obtain the value of $\epsilon$ and $\Omega$ with a maximum likelihood fit. Specifically we simulate an ideal adiabatic ramp of $H_{\text{AB}}$ in the 1ES (i.e a three dimensional Hamiltonian) and choose the values of $\epsilon$ and $\Omega$, such that they maximize the probability to measure our obtained data. We obtain $\epsilon=0.22$ and $\Omega=350 \text{Hz}$.

In the main text we state that tomography of the resulting prepared state is preformed by occupancy and correlation measurements, under assumption of rapid desphasing of other excitation subspaces. Indeed, assuming the prepared state, $\ket{\psi}=\sqrt{p_1}e^{i\phi_1}\ket{100}+\sqrt{p_2}e^{i\phi_2}\ket{010}+\sqrt{p_3}e^{i\phi_3}\ket{001}$, then the occupancy measurement is equivalent to evaluating $\langle\sigma_z\rangle$ for each of the three sites. Similarly, evaluating $\langle\sigma_j\sigma_{j+1}\rangle$ after the in-phase measurement procedure yields $\frac{2}{3}p_{j} p_{j+1}\cos\left(\theta_{j}-\theta_{j+1}\right)$, and after the out-of phase correlation procedure yields, $\langle\sigma_1\sigma_2\rangle=\frac{2}{3}p_1 p_2\sin\left(\theta_1-\theta_2\right)$ and $\langle\sigma_2\sigma_3\rangle=\frac{2}{3}p_2 p_3\sin\left(\theta_2-\theta_3\right)$. These combined are enough to reconstruct $\ket{\psi_f}$ up to a global phase.

\begin{figure}
\centering
	\includegraphics[width=1\linewidth]{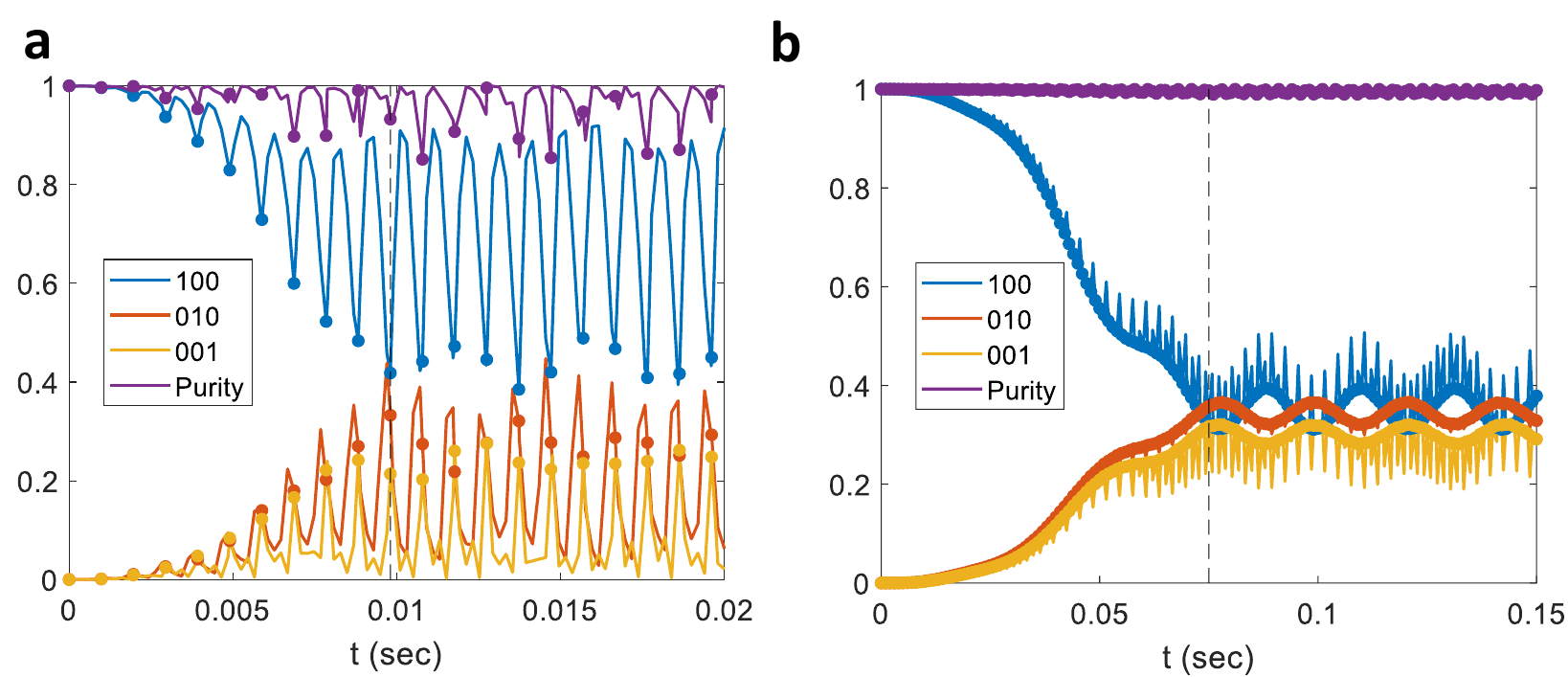}
	\caption{Simulation of adiabatic ground state preparation of the AB ring. (a) Preparation with parameters similar to the experiment, with $\Phi=\pi/2$ and $\beta=7$. The adiabatic ramp starts at $t=0$ and ends at $t=10\text {msec}$ (dashed). The resulting state occupies the three 1ES states almost equally at stroboscopic times (markers), and is coherent, seen as by the high purity values. (b) Preparation deep ion the adiabatic regime with $\beta=40$ and $\Phi=0$, the system evolves smoothly at all times, and results in a high overlap with the AB Hamiltonian ground state.}\label{figRampSim}
\end{figure}

To supplement the data we preform a numerical simulation of the full ion-chain Hamiltonian of the adiabatic ramp (i.e a $2^3 (n_\text{max}+1)$ dimensional Hamiltonian, with $n_\text{max}=9$ the maximum phonon occupation number of the chain's center-of-mass normal mode of motion). Figure \ref{figRampSim}(a) shows the evolution of the different 1ES states under the adiabatic ramp protocol with $\Phi=\pi/2$ and the remaining parameters similar to those of the experiment, such that $\beta=\xi/2\Omega_n=7$ \cite{manovitz2020quantum}. The ramp starts at $t=0$ and ends at  $10\text{ msec}$ (dashed) after which the resulting state is coherent, seen from its high purity (purple) and occupies almost equally the three 1ES states (blue, red and yellow) at the stroboscopic times (markers), indicating an approximate ground state of the underlying AB Hamiltonian. 

We also repeat this simulation deep in the adiabatic regime, with $\beta=40$ and $\Phi=0$, shown in Fig. \ref{figRampSim}(b). Here the procedure yields an even higher overlap with the underlying ground state. Furthermore, the large $\beta$ value results in much less oscillations making the result viable at all evolution times.

\subsection*{Triangular model analysis}
We provide further details on the $4$ site triangular ladder. Specifically we show that the three symmetries, $U_{1,4}$, $A_{2,3}$ and $C$, described in the main text, allow to easily recognize $7$ of the $16$ eigenstates. Furthermore this analysis immediately shows periodicity of the 1ES evolution and of a subspace of the 2ES. A summary of the eigenstate is given in Table \ref{tblEigenstates}.

\begin{table}
\caption{Eigenstate of the 4 site triangular ladder at $\Phi_S=\pi/2$, which are straightforward to obtain. Showing the state, energy and excitation subspace of each eignestate.}
\label{tblEigenstates}
    \begin{center}
                \begin{tabular}{C{3cm} C{2cm} C{2cm}} 
         \hline\hline
         State  & $\langle H\rangle $ & nES \\ [0.5ex] 
         \hline\hline
         $\ket{\downarrow\downarrow}\ket{\downarrow\downarrow}$ & 0 & 0 \\ 
         \hline
         $\ket{\uparrow\uparrow}\ket{\uparrow\uparrow}$  & 0 & 4 \\
         \hline
         $\ket{S}\ket{\downarrow\downarrow}$ & 0 & 1  \\
         \hline
         $\ket{S}\ket{\uparrow\uparrow}$ & 0 & 3  \\
         \hline
         $\ket{S}\frac{\ket{\uparrow\downarrow}-i\ket{\downarrow\uparrow}}{\sqrt{2}}$ & $\Omega$ & 2  \\ 
         \hline
          $\ket{S}\frac{\ket{\uparrow\downarrow}+i\ket{\downarrow\uparrow}}{\sqrt{2}}$ & $-\Omega$ & 2  \\ 
         \hline
         $\frac{\ket{\uparrow\uparrow}\ket{\downarrow\downarrow}-\ket{\downarrow\downarrow}\ket{\uparrow\uparrow}}{\sqrt{2}}$ & 0 & 2 \\[1ex]
         \hline\hline
        \end{tabular}
    \end{center}
\end{table}

We start by noticing that the unoccupied state, $\ket{\downarrow\downarrow}\ket{\downarrow\downarrow}$, and the fully occupied state,  $\ket{\uparrow\uparrow}\ket{\uparrow\uparrow}$ are trivially $E=0$ eigenstates of the system as all Hamiltonian terms annihilate them. Here and in all other states below the two kets are to be understood as $\ket{\psi}\ket{\varphi}\equiv\ket{\psi}_{1,4}\ket{\varphi}_{2,3}$.

Next, due to $U_{1,4}$, when sites $1$ and $4$ form a spin singlet they are decoupled from the evolution. Therefore a set of eignestates can be found by diagonalizing the remaining terms connecting sites $2$ and $3$. In the 1ES this comes about as $\ket{S}\ket{\downarrow\downarrow}$, with $E=0$. The chiral symmetry now forces the three remaining eigenstates of the 1ES to have energies $0,\pm E_1$, with $E_1$ determined by exact diagonalization of the $3\times3$ Hamiltonian. Similarly in the 3ES we have $\ket{S}\ket{\uparrow\uparrow}$, with $E=0$.

In the 2ES a singlet on sites $1$ and $4$ require and additional excitation on sites $2$ and $3$. The interaction between these sites takes the form of a $\sigma_y$ term, thus two eigenstates in the 2ES are  $\ket{S}\frac{\ket{\uparrow\downarrow}\pm i\ket{\downarrow\uparrow}}{\sqrt{2}}$ with energy $\mp\Omega$. The state $\frac{\ket{\uparrow\uparrow}\ket{\downarrow\downarrow}-\ket{\downarrow\downarrow}\ket{\uparrow\uparrow}}{\sqrt{2}}$ is an additional eigenstate in the 2ES, carrying 0 energy. Similarly to above, the remaining three states must carry energies $0,\pm E_2$, with $E_2$ determined by exact diagonalization.

\end{document}